# Programmable reaction-diffusion fronts

Anton S. Zadorin[‡], Yannick Rondelez[∥], Jean-Christophe Galas[‡], and André Estevez-Torres[‡]

‡. Laboratoire de photonique et de nanostructures, CNRS, route de Nozay, 91460 Marcoussis, France.

∥. LIMMS/CNRS-IIS, University of Tokyo, Komaba 4-6-2 Meguro-ku , Tokyo, Japan.

**Corresponding author:** André Estevez-Torres (aestevez@lpn.cnrs.fr)

## Table of contents

**Abstract:**

Morphogenesis is central to biology but remains largely unexplored in chemistry. Reaction-diffusion (RD) mechanisms are, however, essential to understand how shape emerges in the living world. While numerical methods confirm the incredible potential of RD mechanisms to generate patterns, their experimental implementation, despite great efforts, has yet to surpass the paradigm of stationary Turing patterns achieved 25 years ago. The principal reason for our difficulty to synthesize arbitrary concentration patterns from scratch is the lack of fully programmable reaction-diffusion systems. To solve this problem we introduce here a DNA-based system where kinetics and diffusion can be individually tuned. We demonstrate the capability to precisely control reaction-diffusion properties with an autocatalytic network that propagates in a one-dimensional reactor with uniform velocity, typically 100 $\mu$m min$^{-1}$. The diffusion coefficient of the propagating species can be reduced up to a factor 2.7 using a species-specific strategy relying on self-assembled hydrodynamic drags. Our approach is modular as we illustrate by designing three alternative front-generating systems, two of which can pass through each other with little interaction. Importantly, the strategies to control kinetics and diffusion are orthogonal to each other resulting in simple programming rules. Our results can be quantitatively predicted from first-principle RD equations and are in excellent agreement with a generalized Fisher-Kolmogorov-Petrovskii-Piscunov analytical model. Together, these advances open the way for the rational engineering of far-from-equilibrium arbitrary patterns and could lead to the synthesis of self-organizing materials.

**Significance Statement:**

How macroscopic spatiotemporal order arises in a system of chemical reactions is a long-standing question which has important implications in biological morphogenesis. Traveling waves of concentration and stationary Turing patterns, which dynamics are ruled through an interplay between reaction and diffusion, are the archetypes of the emergence of such an order. However, these important phenomena have been interrogated so far with a class of chemical reactions for which reaction and diffusion are hardly tunable: redox or pH oscillators such as the Belousov-Zhabotinsky reaction. Here we report a programmable DNA-based biochemical system where both the reaction and the diffusion terms can be easily controlled. It opens the way to the synthesis of reconfigurable reaction-diffusion behaviors.

**Author contributions:** AET designed research; AZ and AET performed research and analyzed data; JCG and YR contributed new reagents/analytic tools; all the authors discussed the results and wrote the paper.

# 1. Introduction

Morphogenesis is an area that remains largely unexplored in chemistry. We know, however, that reaction-diffusion (RD) mechanisms are essential for the emergence of spatio-temporal ordered structures in living systems (1). Our capacity to generate concentration patterns from scratch hence bears the potential to increase our understanding of morphogenic processes in an unprecedented way. Turing (2) and later Gierer and Meinhardt (3) laid the theoretical foundations of chemical morphogenesis. At steady state, a large majority of chemical systems relax to a state of homogeneous concentration. Many excitable systems and oscillators develop fronts, waves and spirals (4-7), with well-defined velocity. But just a handful of reactions produce more complex behaviors such as stationary Turing patterns (8-10), replicating (11) and oscillating spots (12). Nothing more complex than that has ever been observed experimentally in synthetic systems in the absence of external forcing (13). Computational methods suggest, however, a wealth of possible phenomena (13, 14) but these are difficult to test because experimental systems with tunable properties have remained elusive in the laboratory. In this work we introduce a programmable reaction-diffusion experimental system.

To generate arbitrary spatio-temporal patterns the following properties need to be programmable: i) the topology of the chemical reaction network (CRN), ii) the reaction rates, $r_i$ and iii) the diffusion coefficients of the individual species, $D_i$. The first two requirements guarantee a chemical system with non-trivial dynamics. The last two conditions allow to probe experimentally different regions of the bifurcation diagram. In the last forty years, a strong effort has been dedicated to develop experimental systems where the aforementioned properties could be programmed. The majority of them are related to the Belousov-Zhabotinsky (BZ) reaction (15-17): they concern small inorganic or organic molecules and redox or acid-base reactions (we will call them BZ-related reactions). Our current understanding of chemical reactivity does not allow to engineer CRNs with such chemistries in a rational way. Although semi-heuristic methods have been developed (18-20), they are neither general nor modular. They are nevertheless still the gold standard to experimentally test reaction-diffusion theories (21). An essential point in the quest to synthesize complex RD patterns is the ability to selectively reduce $D_i$ for a given chemical species (1, 2). Particular solutions to reduce $D_i$ have been devised for BZ-related reactions (8, 10) but no general strategy is available.

DNA-based chemical reaction networks provide an interesting solution to the issues mentioned above. Due to base complementarity, the reactivity of single stranded DNA (ssDNA) towards hybridization can be predicted from the sequence (22, 23). Recent advances in DNA nanotechnology allow us to program the topology of quite complex CRNs. Enzyme-free DNA circuits have produced tunable cascading reactions (24, 25) and they have recently allowed to encode edge detection algorithms of light-generated patterns (26). In combination with enzymatic reactions, non-equilibrium dissipative behaviors with DNA circuits have been obtained, such as non-linear oscillators (27-29) and memory switches (30). We have recently observed wave trains and spirals in a synthetic reaction network made of short DNA strands and three enzymes (31). Here we introduce a general method to control specifically both the reaction rates and the diffusion coefficients of DNA species involved in such programmable reaction networks. We apply it to a far-from-equilibrium autocatalytic system that develops traveling fronts of uniform velocity. As a result, we demonstrate that the propagation velocity of the fronts can be controlled by either reaction or diffusion in a predictable manner. Importantly, a source of dNTPs maintains the system far from equilibrium for several hours in a closed reactor, which greatly facilitates its experimental implementation. The growing field of structural DNA nanotechnology (32) uses DNA-encoding to self-assemble $\mu$m structures with nanometer resolution (33, 34). We show that the same DNA-based chemistry can be harnessed to generate order on the millimeter length scale and hence suggest that both approaches could be bridged in future integrative models of the emergence of shapes in the living world.

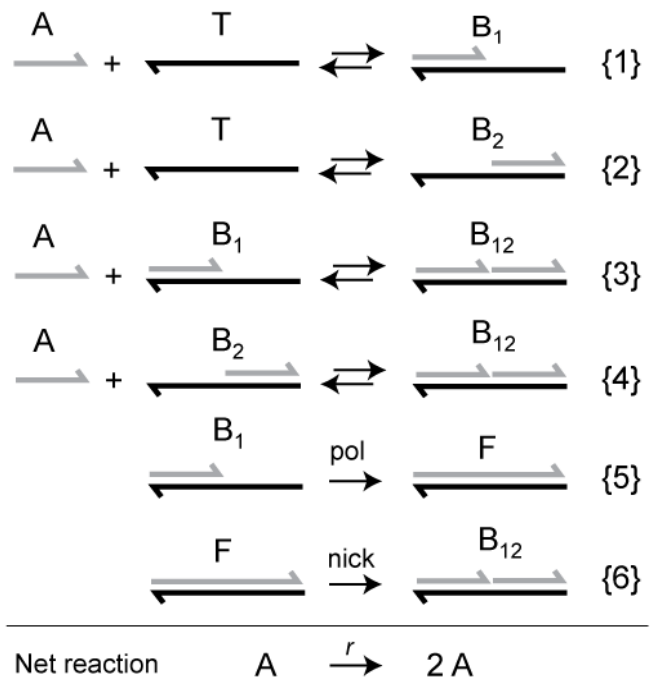

**Figure 1:** Mechanism of the DNA-based autocatalyst. The 11-base long DNA strand A reversibly hybridizes with a 22-mer template, T, bearing two contiguous sites complementary of A (reactions {1-4}). Species $B_1$ is extended by a polymerase, pol, to form the dsDNA F {5}, which bears a recognition site for a nicking enzyme, nick, yielding $B_{12}$ {6}. The net reaction is the autocatalysis of A with rate $r$. Double and single arrows indicate reversible and irreversible reactions, respectively.

## 2. Model

Throughout this work we consider an autocatalytic system based on the PEN DNA toolbox, (27, 35), which is a modular approach to engineer complex chemical reaction networks. Species A, an 11-mer ssDNA, may catalyze its own growth in the presence of a template, T, a 22-mer, carrying two contiguous domains complementary to A (Figure 1). Species A reversibly hybridizes with T on any of these two domains, generating species $B_1$, $B_2$ and $B_{12}$. $B_1$ may be extended by a polymerase, pol, to form species F. F carries a recognition site for a nicking enzyme, nick, such that the upper strand is cut at its midpoint, yielding species $B_{12}$. The net reaction is thus,

$$A \xrightarrow{r} 2A, \qquad (1)$$

where the rate $r(A)$ depends on $A$, the concentration of A (see SI section 1). In a one dimensional reactor the evolution of $A$ depends on time, $t$, and position, $x$, and is described by the reaction-diffusion equation

$$\frac{\partial A}{\partial t} = r(A) + \frac{\partial}{\partial x}\left(D_{\text{eff}}(A)\frac{\partial A}{\partial x}\right), \qquad (2)$$

where we have made explicit that the diffusion coefficient $D_{\text{eff}}(A)$ may depend on the concentration of A. The reason is that we are modeling a reaction-diffusion system that depends on the concentration of at least 6 different species (Figure 1) with the single average species A. A is either free or bound to T and thus the diffusion coefficient of the average species A, depends on the molar fraction of free A, and thus on $A$. We show in the SI section 2.3 that in our case

$$D_{\text{eff}}(0) \simeq \frac{K_{-1}}{2T_0 + K_{-1}} D_A + \frac{2T_0}{2T_0 + K_{-1}} D_T, \qquad (3)$$

where $K_{-1}$ is the dissociation constant of reaction {1} in Figure 1, $T_0$ is the total concentration of T and $D_A$ and $D_T$ are the diffusion coefficients of species A and T, respectively.

For $r(A) = kA(1 - A/C)$ and $D_{\text{eff}}(A) = D$, Eq. 2 takes the form of the well-known Fisher-Kolmogorov-Petrovskii-Piscunov (Fisher-KPP) equation, where $k$ is the replication rate constant and $C$ the carrying capacity (36, 37). This classic equation has traveling wave solutions of the form $A(x,t) = A(x - vt)$, where the velocity $v$ is bounded from below by

$$v = 2\sqrt{r'(0)D_{\text{eff}}(0)}, \qquad (4)$$

$r'$ being the derivative of $r$ and both $r'(A)$ and $D_{\text{eff}}(A)$ are taken at the limit $A = 0$. In the Fisher-KPP model, $r'(0) = k$ and $D_{\text{eff}}(0) = D$. Importantly, in the case of a) constant $D$, b) $r(0) = 0$, c) bounded growth (*i. e.* there exists $A_{\max} > 0$ such that $r(A_{\max}) = 0$), d) $r(A) > 0$, e) $r'(0) > 0$ and f) $r'(A) < r'(0)$ on $(0, A_0)$, $v$ from Eq. 4 corresponds to the single stable asymptotic traveling wave solution, and depends neither on other details of the growth function $r(A)$, nor on the shape of the initial condition (37, 38). In our experimental conditions a), c) and f) are violated: $D_{\text{eff}}$ depends on $A$, the growth is not bounded (though, it saturates at a certain rate) and it accelerates as *A* increases (in some region of concentrations). However, in the SI we demonstrate theoretically, first, that if a traveling wave solution of the generic equation 2 exists, its velocity will also be bounded from below by Eq. 4 (SI 2.1) and, second, that for a Michaelis-Menten type of growth, $r(A) = kA/(K_M + A)$, the solution with *v* from Eq. 4 does exist (SI 2.2) (SI Figure S1).

Note that the scaling in Eq. 4 is, in principle, valid regardless of the expression of $r(A)$, assuming b) and e) hold, and the change of $r'(0)$ is due to a multiplicative factor on $r(A)$. If the rate-law is unknown, $r'(0)$ is simply the exponential growth rate at low *A*. However, when the function $r(A)$ differs from the one of Fisher-KPP, a multiplicative constant greater than unity may appear in the expression of the velocity (for an exactly solvable chemically relevant example we refer to (39)). To take this into account we introduce a corrective factor γ

$$v^{corr} = \gamma \, v^{mod}, \tag{5}$$

where the indexes stand for corrected and model, respectively, and $v^{mod}$ is given by Eq. 4. Throughout this paper we will show that our programmable molecular system is in quantitative agreement with Eqs. 3-5, with γ = 1.3, and is thus a very good candidate to explore experimentally the emergence of complex patterns.

## 3. Results

### 3.1. A front of autocatalyst propagates with uniform velocity

We first studied the growth kinetics and the front propagation dynamics of autocatalyst A (Figure 2). In both cases the concentration of A was indirectly monitored using the non-specific fluorescent DNA binder EvaGreen. The fluorescence quantum yield of EvaGreen increases 6-fold when bound to dsDNA, compared to ssDNA (SI Table S1). The fluorescence intensity measured in the following is thus proportional to a linear combination of the concentration of double stranded species: $B_1$, $B_2$, $B_{12}$ and F. We first introduced in a well-mixed reactor T, A, pol, nick and deoxyribonucleotides (dNTPs) at 38°C. The growth of the fluorescence intensity displays a sigmoidal shape (SI Figure S3). It is not a simple logistic growth: two growth rates were obtained by fitting with a biexponential function, $k_1 = (8.0 \pm 0.8) \cdot 10^{-2}$ min$^{-1}$, and $k_2 = 0.57$ min$^{-1}$ (Figure 2A and SI Figure S3). After about 80 min, the fluorescence intensity stabilizes, most likely when all the free templates T are bound to A either as $B_{12}$, $B_1$, $B_2$ or F. The monoexponential growth at low *A* can be explained by a simple kinetic model where the polymerization reaction {5} is rate-limiting (SI Section 1.2). We did not attempt to interpret the second monoexponential time-scale in this work as it appears to be irrelevant for front propagation dynamics.

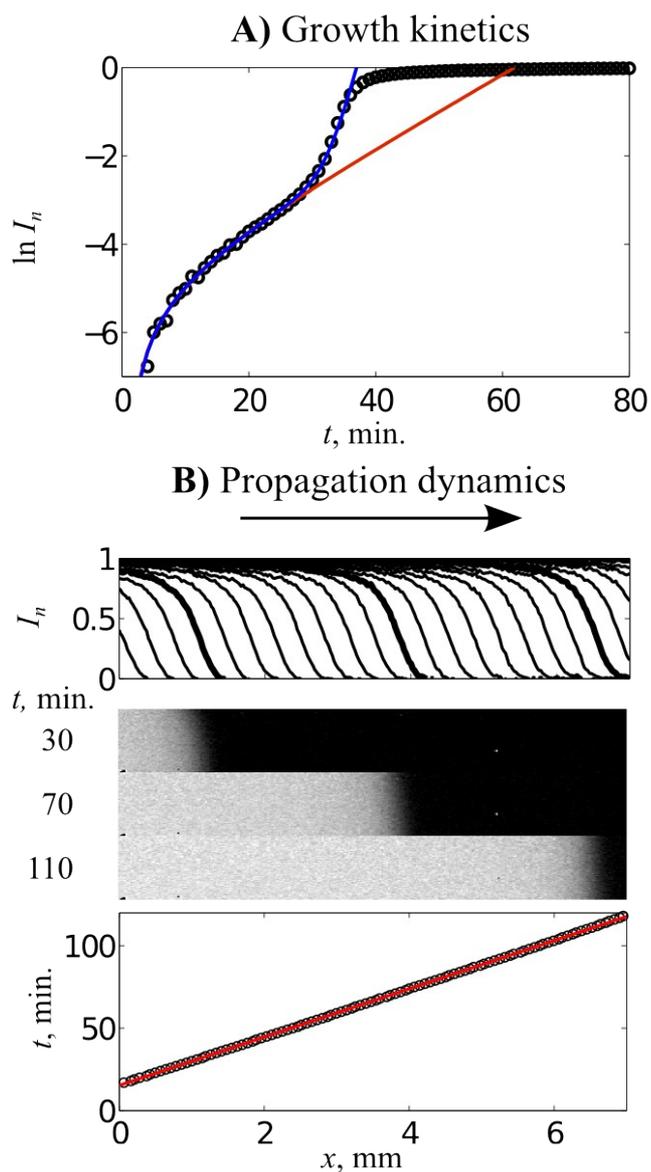

**Figure 2:** The autocatalyst grows exponentially at short times in a well-mixed reactor and generates a front with uniform velocity in a channel reactor. A) Log-lin plot of the normalized EvaGreen fluorescence, $I_n$, vs time, $t$, in a 20 µL tube. The blue line is a biexponential fit between $t = 0$ min and $t = 32$ min. The red line is an exponential fit between $t = 0$ min and $t = 26$ min (see SI Figure S3 for details on the fitting procedure). B) Profiles of $I_n$ along the channel length, $x$, starting from $t = 15$ min in 5 min intervals (top). The arrow shows the direction of the front propagation. The thick lines correspond to the frames shown below. Images of the fronts at 30, 70 and 110 min (middle, SI video S1). Time vs the position of the front (bottom, linear fit in red). $T = 200$ nM, $pol = 16$ U/mL, $nick = 300$ U/mL, 38°C in the reaction buffer with 10 g/L of triton X-100.

When a channel is filled with the reaction buffer with all components except A and an initial condition is created by injection of 1 $\mu$M A to the left inlet, we observe a front of fluorescence that moves from left to right (Figure 2B, SI video S1). The shape of the intensity profile along *x* is stable and propagates with uniform velocity (Figure 2C). The front lasts for about 150 min before reaching the right border of the channel. For 38°C, *T* = 200 nM, *pol* = 16 U/mL, and *nick* = 300 U/mL the velocity of the front is 68 $\mu$m min$^{-1}$. Importantly, and in agreement with the theory described above, the velocity and the shape of the front are independent of the shape and the amplitude of the initial condition (SI Figure S4). Under certain circumstances (higher temperature, higher *pol*) a leak reaction due to the unprimed polymerization of T triggers a homogenous growth across the whole channel, preventing the front from finishing the run. However, in the conditions described above, it takes more than 900 min for the unprimed reaction to become evident (SI Figure S5).

In an independent experiment we measured the diffusion coefficient of fluorescent analogues of A and T at 38°C from the the relaxation of a sharp initial concentration profile (SI section 8, SI Figure S6, Table 2). We obtained $D_A$ = (16 ± 3)·10$^3$ $\mu$m$^2$ min$^{-1}$ and $D_T$ = (10.7 ± 0.7)·10$^3$ $\mu$m$^2$ min$^{-1}$, respectively, in agreement with values reported in the literature (40, 41). As a proxy for $K_{-1}$ we measured the dissociation constant of the hybridization of A with its complementary strand and found 3 nM at 38°C. From Eqs. 3-5 we thus calculate the values predicted by the model: $D_{eff}^{mod}(0)$ ≈ (10.7 ± 0.7)·10$^3$ $\mu$m$^2$ min$^{-1}$, $v^{mod}$ = 59 ± 7 $\mu$m min$^{-1}$ and $\gamma$ = 1.1 ± 0.2. The front velocity predicted by the simple Fisher-KPP model is just 16% below the value measured experimentally.

## 3.2. The velocity of the front depends on the growth rate, which can be specifically tuned

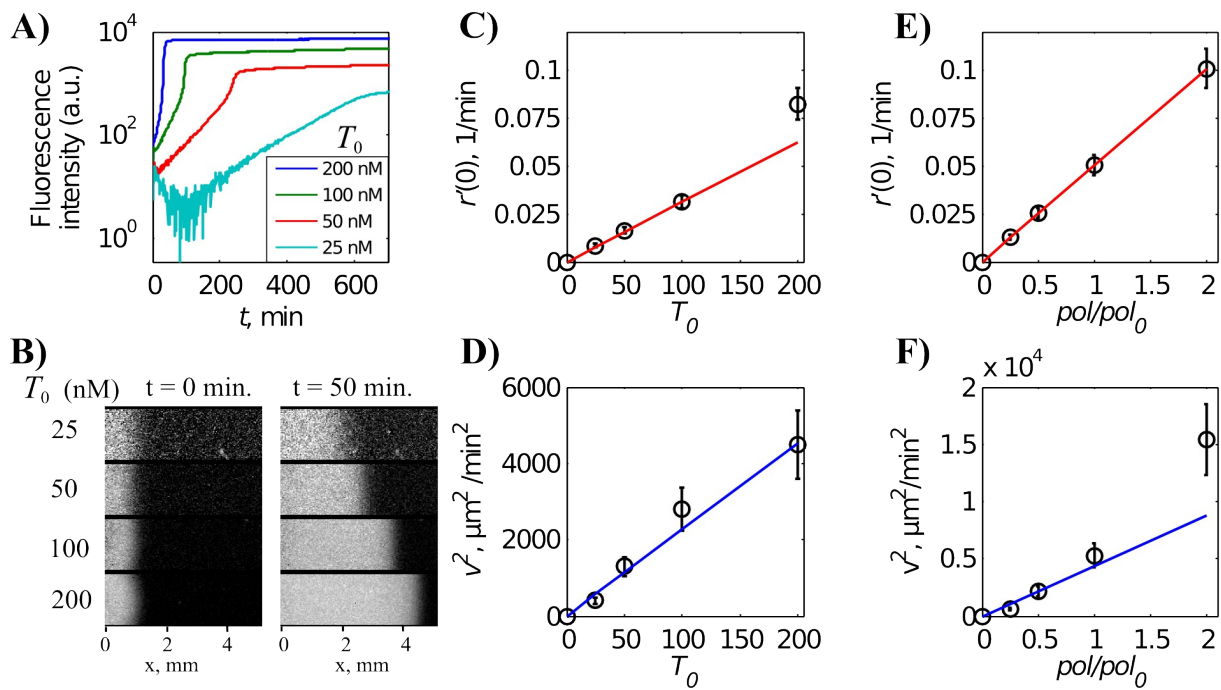

**Figure 3:** The growth rate of the autocatalyst and its propagation velocity can be tuned specifically with the template concentration and non-specifically with the polymerase concentration. A) Log-lin plots of the growth kinetics with different concentrations of the template, $T_0$. B) Fluorescent images of the front position at 0 min and 50 min for different $T_0$. For clarity, the brightness of the images with different $T_0$ has been normalized (SI video S2). C) First order rate constant $r'(0)$ vs $T_0$, the red line is a linear fit for $T_0$ = 0–100 nM. D) Square of the front velocity $v$ vs $T_0$, the blue line is the prediction using Eqs. 3-5 with $\gamma$ = 1.3. E) $r'(0)$ vs normalized polymerase concentration, $pol/pol_0$, the red line is a linear fit. F) $v^2$ vs $pol/pol_0$, the blue line is the prediction using Eqs. 3-5 with $\gamma$ = 1.3 (SI video S3). Experimental conditions: A-D) 38°C, $pol$ = 16 U/mL, $nick$ = 300 U/mL, E-F) $T_0$ = 200 nM, $pol_0$ = 16 U/mL, $nick$ = 500 U/mL, 44°C. Error bars were estimated from the 10% experimental precision (both on $r'(0)$ and $v$) measured for 4 independent experiments at $T_0$ = 200 nM (Table 1-2).

We went out to check both the scaling $v \sim (r'(0))^{1/2}$ (Eq. 4) and the capability of our model to provide a quantitative prediction of $v$ with a unique value of $\gamma$. To do so we performed growth and front propagation experiments with different concentrations of T and pol (Figure 3). Figure 3A-D reports the dependence of growth and propagation on the total template concentration, $T_0$. The growth of the autocatalyst is always monoexponential at short times for $T_0$ = 0–200 nM. This monoexponential character is maintained during the whole growth phase at 25 nM, but a second exponential time-scale appears at larger $T_0$. For all the values of $T_0$ investigated, the break in the slope in Figure 3A happens at a similar value of intensity, 600 a. u., suggesting that there is a threshold concentration of dsDNA species responsible for a change in the growth mechanism, which becomes faster. This change in the mechanism could be due to the inhibition of the polymerase by the nicking enzyme, which has already been described (31), or to an effect of the single stranded DNA binding protein present in the reaction mix. This fact was not investigated further because the front velocity is set by the growth rate at low $A$. Interestingly, the growth rate is proportional to $T_0$ in the range 0-100 nM, with $r'(0) = (3.1 \cdot 10^{-4}\ \text{nM}^{-1}\text{min}^{-1})T_0$, indicating that, in this range, growth kinetics can be specifically tuned by changing the concentration of their templates, an important feature for modular programmability (Figure 3C). The velocity of the front also depended on $T_0$. Fronts starting at the same position propagated farther within a given time when $T_0$ increased (Figure 3B). As we found $r'(0) \sim T_0$, Eq. 4 predicts $v^2 \sim T_0$, which was verified experimentally for $T_0 = 0\text{-}200$ nM. To quantitatively test the predictions of our model we substituted these data into Eq. 5 and used the independently measured values of $D_i$, $K_{-1}$ and $r'(0)$ to calculate $v^{mod}$ with Eq. 3-4 and then obtain $\gamma = 1.30 \pm 0.16$, in agreement with the value reported above. With this value of $\gamma$, the analytical equation 5 is in excellent quantitative agreement with the data (Figure 3D, blue line).

$T_0$ is thus a convenient experimental parameter to tune the growth rate and the propagation velocity. It has the advantage of being specific: in a complex reaction network with several autocatalysts growing on different templates, changing the template concentration of one of them will modify $r'(0)$ and $v$ for a single autocatalyst. However, it is also desirable to have another experimental knob to set the overall strength of growth and propagation. To this end we studied the dependence of $r'(0)$ and $v$ on the polymerase concentration, $pol$ (Figure 3E-F, SI Figure S7). As a test of the robustness of the model's predictions, we performed these experiments at a different temperature and nicking enzyme concentration, 44°C and 500 U/mL, respectively. For relative pol concentrations ranging from 0.25- to 2-fold the measured values of $r'(0)$ are linearly dependent on $pol$, with $r'(0) = (0.05\ \text{min}^{-1})\ pol/pol_0$. This indicates that in these conditions the polymerization (reaction {5} on Figure 1) is the rate-limiting step. For the same range of $pol$ we measured front velocities between 20 and 125 $\mu$m min$^{-1}$. In an independent experiment we obtained $D_A = (18 \pm 3) \cdot 10^3$ $\mu$m$^2$ min$^{-1}$, $D_T = (11.8 \pm 0.8) \cdot 10^3$ $\mu$m$^2$ min$^{-1}$ and $K_{-1} = 100$ nM at 44°C. In the range $pol/pol_0 = 0\text{-}1$, the velocities predicted by Eqs. 3-5 with γ = 1.30 are in excellent agreement with the experimental ones (Figure 3F, blue line).

## 3.3. Orthogonal autocatalysts with different sequences propagate with uniform but different velocity

To illustrate the versatility of our approach we designed two more autocatalysts $A_1$ and $A_2$ produced by templates $T_1$ and $T_2$, respectively. $A_1$ has 5 out of 11 bases different from A but functions with the same nicking enzyme (Nt.Bst NBI). $A_2$ has 11 out of 11 bases different from A and depends on a different nicking enzyme (Nb.BsmI). All these three species display sigmoidal growth curves with a clear exponential term at the onset of growth (SI Figure S8) with $r'(0)$ for A and $A_2$ is 0.08 min$^{-1}$ and 0.13 min$^{-1}$. The growth of $A_1$ was too fast to allow confident measurement of $r'(0)$. All three autocatalysts developed propagating fronts with a stable shape (Figure 4A) and a uniform velocity (Figure 4B). The propagation velocities of $A_1$ and $A_2$ were 101 and 64 $\mu$m min$^{-1}$, respectively, compared with 70 $\mu$m min$^{-1}$ for A. Using the measured values for $r'(0)$ and identical $D_{eff}(0)$ the corrected calculated velocities with γ = 1.3 are 77 $\mu$m min$^{-1}$ and 97 $\mu$m min$^{-1}$ for A and $A_2$, respectively. The predicted value is in good agreement with experiment for A but not so much with $A_2$ (50% off), suggesting that the factor γ could depend on the template. Indeed, we hypothesize that γ depends on how the full autocatalytic mechanism (Figure 1) is reduced into the single variable reaction-diffusion equation 2.

In a channel containing both T and $T_2$, two fronts propagating in opposite directions could be triggered by injecting A and $A_2$ on the left and right inlet, respectively (Figure 4C and D). For $t$ < 74 min each front propagates in a fresh medium and it is thus not surprising that we observe the same behavior as for independent fronts. After collision, $A_2$ maintains its velocity constant and equal to 66 $\mu$m min$^{-1}$ while the velocity of A is reduced 1.3-fold from 58 to 46 $\mu$m min$^{-1}$. Considering that after collision A and $A_2$ propagate in a region that has, respectively, a high concentration of $A_2$ and A, and thus the potential to saturate the enzymes, the negligible interaction between the two fronts is particularly striking. Here we used, on purpose, two autocatalysts that rely on different nicking enzymes but depend on the same polymerase. Because the substrate concentration is well below the $K_M$ of the polymerase (SI Figure S10) the interaction of the two fronts is negligible. Although we were not able to observe stable colliding fronts when two templates depending on the same nicking enzyme were used, we believe that this technical issue should be solved with a careful optimization of the experimental conditions. In any case, up to eight nicking enzymes with orthogonal recognition sites are commercially available from major manufacturers, which could significantly extend the complexity of the CRNs that can be constructed within the framework of the PEN DNA toolbox. To the best of our knowledge this is the first time that the collision of two chemically distinct fronts is observed. The modularity of the PEN DNA toolbox hence allows to simply design *de novo* autocatalysts that generate predictable and complex spatio-temporal patterns.

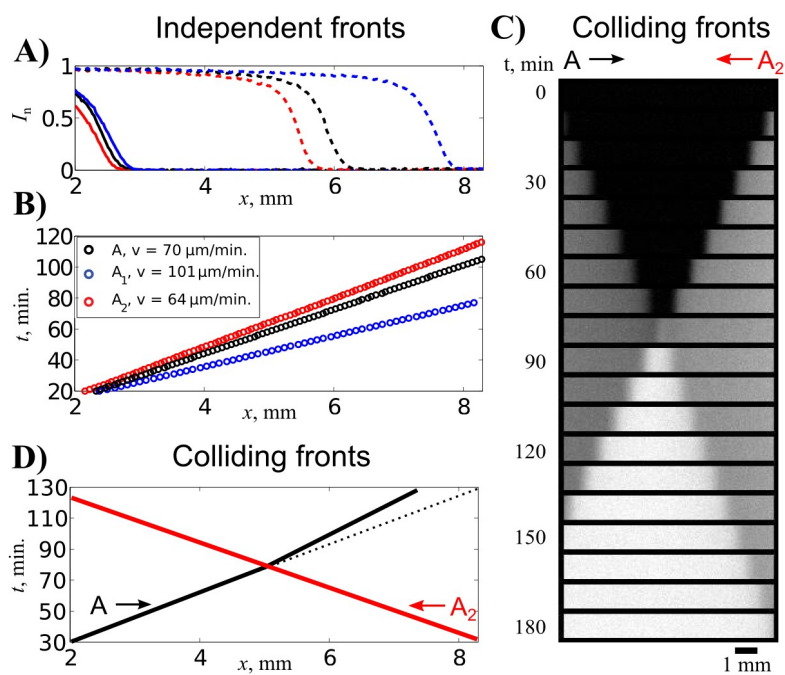

**Figure 4:** Different autocatalysts propagate with different velocities and collide with little interaction. A and $A_1$ differ by 5 bases and depend on the same nicking enzyme, Nt.Bst NBI, while $A_2$ uses another nicking enzyme, Nb.BsmI. A) Profile of the fronts generated by different autocatalysts in separate channels at $t = 20$ min (solid lines) and at $t = 70$ min (dashed). B) Time *vs* the position of the front for data in panel A. C) Time-lapse images (SI video S4) and D) time *vs* the position of the front for a front of A, propagating left to right, and a front of $A_2$, propagating right to left, colliding at $t = 74$ min and $x = 5$ mm. The dotted line in panel D is a guide to the eye to appreciate the slope-break after collision. The color code is conserved within the figure with A in black, $A_1$ in blue and $A_2$ in red. $T = T_1 = T_2 = 200$ nM, pol = 16 U/mL, nick = 300 U/mL, 38°C.

## 3.4. The diffusion coefficient of an autocatalyst can be selectively reduced with a self-assembled hydrodynamic drag

So far we have shown three independent strategies to modify the growth rate and thus the propagation velocity. We develop in the following a method to reduce $D_{eff}(0)$ without modifying the growth rate and thus changing the front velocity through diffusion (Eq. 4). Controlling the diffusion coefficient, $D$, of a molecule is not a simple task. Indeed, $D \sim R^{-1}$, $R$ being the hydrodynamic radius of the molecule, but $R \sim M^{1/2}$, where $M$ is the molecular mass, in the case of a random coil. As a result, relatively large monomolecular entities need to be involved if one wants to reduce $D$ significantly. However, these entities need not necessarily be covalent or even stable: if A interacts dynamically with a ligand, its effective diffusion coefficient $D_{eff}(0)$ will be a weighted average between the free state with high $D$ and the bound state with low $D$, as illustrated in Eq. 3. This approach applies well to single stranded DNA species, for which a binding partner always exist as its Watson-Crick complementary. The task then breaks down to reducing the diffusion of that partner.

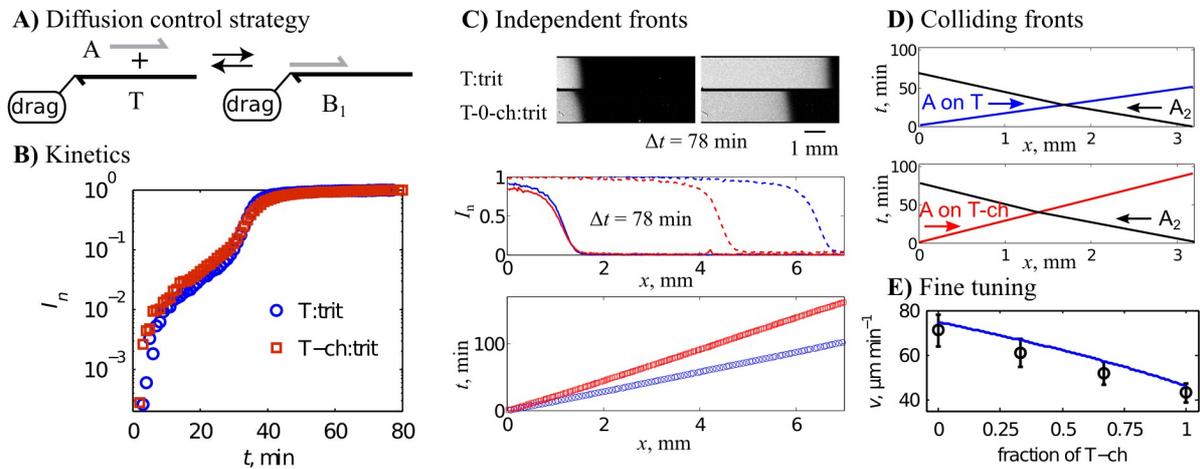

**Figure 5:** The velocity of a front can be reduced using a hydrodynamic drag without altering the growth rate. A) Sketch of the diffusion control strategy implemented in this work, the template, in black, is attached to a hydrodynamic drag, while the active species, in grey, reversibly hybridizes to it. B) Growth kinetics of normalized fluorescence *vs* shifted time. C) Propagation of fronts generated by templates T:trit (blue curves) and T-ch:trit (red curves) in different channels, represented as still images at two given times (top, SI video S5), as fluorescence profiles along the channel at the same times (middle, solid lines t=0 min, dashed t=78 min), and as time *vs* position of the front at half height (bottom). D) Time *vs* front position for colliding fronts of A on T and $A_2$ (top) and A on T-ch and $A_2$ (bottom). E) Fine-tuning of the front velocity through diffusion by changing the molar fraction of T-ch compared to T and keeping constant the total concentration of template, the line is the theoretical prediction from Eqs. 3-5 and $\gamma = 1.30$. Template concentration 200 nM (except in panel D, 150 nM), *pol* = 16 U/mL, *nick* = 300 U/mL, $A_0$ = 10 nM (growth kinetics), 10 g/L triton X-100, 38°C.

The strategy used here consists of attaching a hydrodynamic drag to the 3'-end of template T, which binds to the active species A (Figure 5A). We considered two types of drags: permanently and dynamically attached ones. Permanent drags may be bound to a DNA strand through an irreversible interaction, such as streptavidin-biotin (this work), acrydite-acrylamide (42), or amide-coupling (43). They require, however, a preliminary and cumbersome coupling process. In contrast, dynamic drags reversibly bind to a DNA strand, for instance through hydrophobic interactions. They do not need a coupling step because they self-assemble in solution. We have tested two permanent drags, streptavidin and streptavidin-coated beads, and one dynamic drag, micelles. The micelles were made of triton X-100, a neutral surfactant. In the first case, T was linked in 3' to a biotin through a 5 thymine spacer, which is noted T-5-bt, and further coupled to streptavidin to obtain T-5-bt:str. In the second case, T was linked in 3' to a cholesteryl without thymine spacer, noted T-ch, and used in a 10 g/L triton solution yielding T-ch:trit. At this concentration triton self-organize into micelles about 5.5 nm in radius and the cholesteryl group dynamically attaches to them through hydrophobic interactions. Table 1 and 2 provide sizes for these drags and for species A, T, T-5-bt, T-ch, as well as their corresponding diffusion coefficients at 44°C or 38°C. Triton micelles worked best and are described in the following. Streptavidin worked well (SI Figure S12) but resulted in a reduction of $D_{eff}(0)$ of only 1.6-fold. Beads strongly reduced the diffusion coefficient of T but they did not provide a control strategy orthogonal to growth kinetics (SI Figure S13).

**Table 1:** Estimated hydrodynamic radius, $R_h$, measured diffusion coefficient, $D$, growth rate, $r'(0)$, front velocity, $v$, and an inferred diffusion coefficient associated to the front propagation, $D_{eff}(0)$. Template concentration 200 nM, $pol$ = 16 U/mL, $nick$ = 500 U/mL. All measurements were performed at 44°C, except for $D$ that was done at 20°C or 38°C and recalculated to 44°C. Where applicable, values are accompanied by confidence intervals with the confidence probability of 0.95. The intervals are calculated for samples of n = 3 for $r'(0)$ and n = 4 for $v$, $D_{eff}(0)$ was treated as a function of these variables.

| Species i | $R_h$ (nm) | $D_i$ ($10^3$ $\mu m^2$ $min^{-1}$) | $r'(0)$ ($10^{-2}$ $min^{-1}$) | $v$ ($\mu m$ $min^{-1}$) | $D_{eff}(0)$ ($10^3$ $\mu m^2$ $min^{-1}$) |
|---|---|---|---|---|---|
| A | - | 18 ± 3[3] | - | - | - |
| T | 1.5[1] | 11.8 ± 0.8[3] | 5.6 | 73 | 24 |
| T-5-bt | 1.5[1] | 13.6 ± 0.8[4] | 6.6 ± 0.2 | 87 ± 10 | 29 ± 7 |
| T-5-bt:str | 1.6[2] | 7.6 ± 0.5[4] | 6.2 ± 0.1 | 66 ± 6 | 18 ± 3 |

[1]Calculated from (40). [2]Size of streptavidin alone, from (44). [3]Recalculated from measurements at 38°C. [4]Recalculated from measurements at 20°C.

**Table 2:** Estimated hydrodynamic radius, $R_h$, measured diffusion coefficient, $D$, growth rate, $r'(0)$, front velocity, $v$, and an inferred diffusion coefficient associated to the front propagation, $D_{eff}(0)$. Template concentration 200 nM, *pol* = 16 U/mL, *nick* = 300 U/mL, 10 g/L triton X-100. All measurements were performed at 38°C. Where applicable, values are accompanied by the confidence interval with the confidence probability of 0.95. The intervals are calculated for samples of n = 5 both for $r'(0)$ and $v$, $D_{eff}(0)$ was treated as a function of these variables.

| Species | $R_h$ (nm) | $D_i$ ($10^3$ $\mu m^2$ min$^{-1}$) | $r'(0)$ ($10^{-2}$ min$^{-1}$) | $v$ ($\mu m$ min$^{-1}$) | $D_{eff}(0)$ ($10^3$ $\mu m^2$ min$^{-1}$) |
|---|---|---|---|---|---|
| A | - | 16 ± 3 | - | - | - |
| T:trit | 1.5[1] | 10.7 ± 0.7[3] | 7.7 ± 1.3 | 65 ± 5 | 14 ± 3 |
| T-ch:trit | 5.5[2] | 4.0 ± 0.3[4] | 7.8 ± 0.5 | 40 ± 4 | 5.1 ± 1.1 |

[1]Calculated from (40). [2]Size of triton micelles alone at 30°C, from (45). [3]Value for T in a triton-free buffer. [4]Compatible with the value 2.5·$10^3$ $\mu m^2$ min$^{-1}$ measured for triton micelles alone at 30°C (45).

We first checked the influence of the triton drag on the growth kinetics. Figure 5B displays the growth curves for T and T-ch in the presence of 10 g/L triton X-100. Both curves display a biexponential and remarkably similar shape, with equal growth rates within experimental error (Table 2). They differ only by a shift of 18 min in the onset of growth, which is negligible taking into account the experiment-to-experiment variation (SI Figure S9). Furthermore, control experiments demonstrate that for both templates the polymerization rates were identical, while nicking rates differed by a factor 3 (SI Figures S10 and S11). Considering that polymerization is the rate-limiting step these data demonstrate that the triton drag strategy has a negligible influence in the growth kinetics. Figure 5C shows the propagation of a front of A growing on either T or T-ch in a triton solution with the same reaction conditions. The second front advances 1.6 ± 0.2 times slower, the velocities being 65 ± 5 and 40 ± 4 $\mu m$ min$^{-1}$, respectively (confidence 0.95). These values, together with the growth rate, can be substituted into Eq. 4, yielding a (2.7 ± 0.8)-fold reduction in $D_{eff}(0)$. To compare these with the prediction given by Eqs. 3-5, we independently measured the diffusion coefficient of T-ch:trit, $D_{\text{T-ch:trit}}$ = (4.0 ± 0.3)·$10^3$ $\mu m^2$ min$^{-1}$. Supposing that the hybridization constant is not affected by the presence of triton and taking thus $K_{-1}$ = 3 nM, together with γ = 1.30, the predicted velocity for the front growing on T-ch is 46 ± 2 $\mu m$ min$^{-1}$. Moreover, the ratio

$$\frac{D_{eff}^{no-drag}(0)}{D_{eff}^{drag}(0)} = \frac{K_{-1}D_A + 2T_0 D_T}{K_{-1}D_A + 2T_0 D_{T:drag}}$$

can be used to estimate the theoretical expectation of the change of $D_{eff}(0)$ in the presence of a drag. For T-5-bt and T-5-bt:str at 44°C from Table 1 we obtain a (1.5 ± 0.1)-fold change, while for T and T-ch:trit at 38°C from Table 2 this ratio is 2.6 ± 0.2. All the predicted values are thus in excellent agreement with the experimental figures indicating that the diffusion coefficient of a propagating autocatalyst can be tuned in a quantitative manner.

We further demonstrated that this diffusion control strategy worked well when fronts of A and $A_2$ collided in channels containing either T or T-ch and $T_2$ (Figure 5D). The velocity of A growing on either T or T-ch was not influenced by the front of $A_2$ and again a velocity reduction factor of 1.7 was measured. The presence of the drag on another template did not influence at all the velocity of $A_2$. This shows that diffusion control can be performed selectively on a single node of a chemical reaction network and suggests that it may be scaled to larger networks. Finally, in a channel containing both T and T-ch, we achieved fine tuning of the velocity of a front of A by varying the molar fraction of T-ch while keeping the total concentration (*T+T-ch*) constant (Figure 5E). This fine tuning was exclusively due to diffusion control. The analytical prediction with γ = 1.3 is, once again, in excellent agreement with the experimental data, without fitting.

## 4. Discussion

The first to suggest a connection between a reaction-diffusion process and the morphology of an organism was Alan Turing in 1952 (2). He demonstrated that two chemicals that react and diffuse may create an inhomogeneous stationary spatial pattern of well-defined wavelength from a homogeneous initial condition. A key constrain for this to happen is that the diffusion coefficient of the activator species, the autocatalyst, needs to be significantly smaller than that of the inhibitor (46). Although this was a foundational work with great impact, it took nearly forty years for chemists to get experimental evidence of Turing patterns (8, 9). The reason is that chemical systems used so far to investigate dissipative structures have involved small inorganic and organic molecules, for which the reaction rates, the mechanism and the diffusion coefficients can be hardly modified in a rational manner. Although Epstein (15) and others have made extraordinary experimental and theoretical contributions to the understanding of dissipative chemical structures, the field has reached an *impasse* from the experimental point of view because of the lack of powerful control tools.

Recently, purified biochemical models have been used to study striking spatio-temporal phenomena, in particular the Min system (47). Such systems have the advantage of being biologically relevant, but remain hard to reprogram. We argue here that DNA-based biochemical systems are experimental models of choice to study the emergence of spatio-temporal order in chemistry with important implications in both biological morphogenesis and in the synthesis of self-organizing materials. We think that they will advantageously replace Belousov-Zhabotinsky-related systems. In a previous work we have shown that relatively complex chemical reaction networks (CRNs) can be designed from the bottom up into DNA-based biochemical systems (29) and they display traveling waves and spirals in a non-stirred reactor (31). Here we further demonstrate that the velocity of traveling fronts in a related autocatalytic system can be finely and quantitatively controlled. This control arises from the modularity of the DNA-toolbox and from the specificity of the biochemical reactions involved. Our system has three types of chemical species: active species, A, templates, T, and enzymes, pol and nick. The total concentration of active species changes over space and time; they can be generated, or degraded and they diffuse driven by large gradients. In contrast, the total concentration of templates and enzymes does not change over time or space. The rate constants of a given CRN depend mainly on the total concentrations of templates and enzymes and only to the second order (when saturation arises) on the concentration of free species. As a result, the rate constants for each reaction can be set independently by changing the concentration of a polymerase, the concentration of a template or its sequence. This is impossible to do for BZ-related systems in a closed reactor. To overcome this problem, cumbersome open reactors in contact with the top and bottom of a thin sheet reactor, were needed to observe complex spatio-temporal structures in BZ-related systems (8, 20, 48). Such reactors guarantee that the concentration of a given chemical (and thus the rate of each reaction) is constant over time. They are not needed in systems designed on the framework of the PEN DNA toolbox (27, 35) due to the characteristics of enzymatic reactions: for a chemical B reacting with an enzyme with Michaelis-Menten constant $K_M$, the rate of the reaction is constant and independent of $B$ for $B \gg K_M$. This happens in our case for dNTPs and polymerase, for instance: the excess of dNTPs acts as a reservoir of free energy keeping polymerization rate constant over long periods of time (100-1000 min).

We also have shown that the programmability of DNA and its chemical versatility (many chemical modifications are commercially available) make it straightforward to design selective strategies to control the diffusion coefficient of an active species. Two strategies to reduce the diffusion coefficient have been used in the past in BZ-related systems. The medium was supplemented with starch, that made a complex with iodide (8), or the reaction was carried out in a water-in-oil emulsion (10). In this last case bromide could diffuse rapidly from one water droplet to another, as it is soluble in oil, but the hydrophilic activating species could only move from one droplet to another through droplet merging, which is a slow process. These two implementations depend on the intrinsic properties of the reactants and are neither general nor modular, in contrast with the strategy shown here. Moreover we have demonstrated that our strategies to control kinetics and diffusion are orthogonal to each other; when diffusion is modified kinetics is not, making our system easily programmable. It has, of course, limitations. The presence of triton in the solution facilitates the formation of bubbles, specially at 38 ºC, which may cause trouble. Moreover, the diffusion control needs a low value of the dissociation constant between A and T (Eq. 3) while the selective control of the growth rate using the template concentration needs the opposite. Finally, while the presence of cholesteryl and triton had a marginal effect in the overall growth rate it did influence the rate of the nicking reaction.

Importantly, the biochemical system presented here is commercially available, relatively cheap, very robust to the variability of enzymatic activity inherent to commercial preparations, simple to carry out and compatible with widely available reactor materials such as polystyrene. It does not require particular skills in biochemistry: no need for protein or DNA purification, for instance. The spatial reactor was fabricated with low-tech protocols using plastic slides and Parafilm (SI Figure S14) available in any laboratory. For these reasons, we anticipate that DNA-based systems will be a widely used experimental model to ask fascinating questions about the emergence of spatio-temporal molecular order (49).

## 5. Conclusion

We have shown that using a relatively simple chemical system based on DNA and two enzymes it is possible to generate programmable fronts propagating with constant velocity. These fronts can be effectively described by a reaction-diffusion equation with one dependent variable, closely related to the Fisher-KPP problem. This model provides excellent quantitative predictions of the front velocity and its associated effective diffusion with a single phenomenological parameter. We have demonstrated the control of the velocity of the waves *via* kinetics and diffusion. The former can be tuned non-specifically, through the enzyme concentration, or specifically, trough the concentration of a specific DNA template or through its sequence. In addition, we demonstrated a method to control the diffusion coefficient of a DNA reactant by reversible attachment of a self-assembled hydrodynamic drag. Importantly, the methods to control kinetics and diffusion are orthogonal to each other making programming rules simple. The targeted control of diffusion coupled to the simplicity of rewiring a reaction network opens new avenues for the bottom-up construction of fully reconfigurable spatio-temporal dissipative structures.

# 6. Materials and Methods

## 6.1. Reaction assembly

The standard reaction buffer contained 20 mM Tris-HCl, 10 mM $(NH_4)_2SO_4$, 50 mM NaCl, 6 mM $MgCl_2$, 2 mM $MgSO_4$, 10 mM KCl, 0.4 mM of each of dNTP (NEB), 1 g/L synperonic F108 (Sigma Aldrich), 4 mM dithiothreitol, 500 mg/L BSA (NEB), 1 µM netropsin (Sigma Aldrich), 5 mg/L extremely thermostable ssDNA binding protein (ET SSB) (NEB), and 1x EvaGreen DNA binder (20x dilution of the manufacturer's stock solution) (Biotium). When needed, triton X-100 was added to this buffer to the final concentration of 10 g/L to generate micelles. The following two enzymes were added into the mix: Bst DNA polymerase large fragment (pol) (NEB) and Nt.BstNBI nickase (nick) (NEB). In experiments with the $T_2$ template, Nt.BstNBI was substituted by Nb.BsmI (NEB). Typical concentrations were 16 U/mL for pol and 300 U/mL for nick, however, nicking enzyme activities significantly changed from batch to batch, and their concentrations were adjusted according to independent assays. Oligonucleotide sequences are provided in the SI, section 16.

## 6.2. Growth kinetics experiments

Autocatalyst growth independent of spatial variables was achieved by mixing 20 µL of the above solution with 1 µL of 200 nM A, $A_1$, or $A_2$, depending on the template used, (thus, $A_0 \approx 10$ nM). This well-mixed solution was monitored in a CFX96 Touch Real-Time PCR Detection System (Bio-Rad) used as a thermostated fluorescence reader. Alternatively, this reaction mix was injected into a polystyrene chip and monitored under a microscope as described below. Typically, one reaction per experiment was performed in a tube without addition of an autocatalyst (A, $A_1$, or $A_2$) to monitor the onset time of the unprimed growth. Control experiments demonstrate that growth rate constants measured in tubes in the rtPCR machine or in the polystyrene channels used for front propagation were identical within experimental precision. The former method was used for convenience (SI, Figure S15).

### 6.3. Front propagation experiments

The reaction chamber was a channel of approximately 1.8 mm length, 2-3 mm width and 0.25 mm height cut out from two layers of Parafilm and placed between two clear polystyrene slides manually produced from 10 cm Petri dishes. The channel was open on the side from one end, and closed from the other. A hole of 1 mm diameter was drilled in the upper slide above the second end to facilitate the channel filling by aspiration with a micropipette (SI Figure S9). Polystyrene was selected instead of glass because we noticed a strong interaction of the Nt.BstNBI nickase with glass. In contrast, for Nb.BsmI a simple assembly using glass cover slips with no drilling and a channel opened from both sides may be utilized. The Parafilm layers were placed between the slides and left on a hot plate at 50 °C to glue the assembly. The reaction mix with the template but without an autocatalyst was then introduced from the side inlet. To generate the initial condition for the traveling wave, 5 μL of the initial mix were mixed with 0.5 μL of 10 μM A, $A_1$ or $A_2$, then 1.5 μL of the resulting solution was injected from the side. Both the side inlet and the vertical holes were sealed with vacuum grease, and the reaction was then monitored with a Zeiss Axio Observer Z1 inverted microscope with a transparent heating plate (Tokai-Hit) using a 2.5x objective, a HXP 120 C (Zeiss) (experiments in Table 1) or LED (CoolLED) excitation light (all other experiments), a motorized stage with Tango controller (Marzhauser-Wetzlar), and an EM-CCD Digital camera C9100 (Hamamatsu). Images were acquired automatically using μManager 1.4 (50) and treated with ImageJ (NIH). Prior to data analysis, the background and the inhomogeneous illumination were corrected by subtracting the first image and dividing by an average of images where the channel was homogeneously filled with dye.

## 7. Acknowledgments

This work was funded by ANR jeunes chercheurs program under award Dynano. A. Z. acknowledges a postdoctoral fellowship from NanoSciences Ile-de-France (Enginets award) and from the Laboratoire d'Excellence en Nanoscience et Nanotechnologie Nanosaclay (Turnano award). We thank A. Kalley and M. Cabon for preliminary experiments. We are indebted to L. Jullien, C. Gosse and T. Le Saux (ENS, Paris) with whom the idea of diffusion control arouse and to D. Baigl (ENS, Paris) for fruitful discussions. Streptavidin-coated beads were a kind gift from M. Coppey (Curie, Paris).